\documentclass[11pt]{article}

\usepackage[final]{acl}

\usepackage{times}
\usepackage{latexsym}

\usepackage[T1]{fontenc}

\usepackage[utf8]{inputenc}

\usepackage{microtype}

\usepackage{inconsolata}

\usepackage{graphicx}

\usepackage{algorithm}
\usepackage{algorithmic}
\usepackage{amsmath}
\usepackage{amssymb}
\usepackage{booktabs}

\usepackage{xcolor}

\newcommand{\eg}{$\textit{e.g.}$,\ }

\newcommand{\nf}[1]{#1}
\usepackage{multirow}

\usepackage{newfloat}
\usepackage{listings}
\usepackage{bm}

\title{FC-TTS: Style and Timbre Control in Zero-Shot Text-to-Speech\\
with Disentangled Speech Representations}

\author{
Yoonhyung Lee \and Hyunsin Park \and Jinhwan Park \and Jinkyu Lee\thanks{indicates the corresponding author.} \\
Qualcomm AI Research\thanks{Qualcomm AI Research is an initiative of Qualcomm Technologies, Inc.} \\
\texttt{\{yoonhyun, hyunsinp, jinhpark, jinkyu\}@qti.qualcomm.com}
}

\begin{document}
\maketitle

\begin{abstract}
Recent advances in zero-shot text-to-speech (TTS) have enabled accurate imitation of reference speech in terms of both speaking style and speaker timbre. However, achieving disentangled control over these aspects from separate references remains a challenging task. Several studies have proposed disentangled speech representations that decompose speech into interpretable attributes ($\textit{e.g.}$, timbre, prosody, and content), providing a promising foundation for TTS with attribute control from separate references. Yet, how to effectively integrate such representations into TTS systems to achieve independent and precise control remains underexplored. In this paper, we present FC-TTS, a zero-shot TTS framework that enables disentangled control of style and timbre by conditioning on two distinct reference utterances. Unlike existing systems that inherit limitations from those pre-trained disentangled representations, FC-TTS introduces key design strategies, including architectural choices, training framework, and auxiliary training objectives, which improve the reliability of attribute separation and dual-reference control. Experiments show that FC-TTS achieves high-fidelity synthesis and competitive zero-shot naturalness, while uniquely supporting consistent and independent manipulation of style and timbre. Audio samples are available at \url{https://qualcomm-ai-research.github.io/fc-tts}
\end{abstract}

\section{Introduction}
With recent breakthroughs in text-to-speech (TTS) technology~\citep{kim2021conditional, wang2023valle, chen-etal-2024-f5tts}, the focus has expanded from simply generating natural-sounding speech to enabling expressive and personalized synthesis for diverse applications such as virtual assistants, audiobooks, accessibility tools, and interactive media~\citep{Xie2024TowardsCS}.
This evolution has brought a growing demand for fine-grained control over attributes such as speaking style and speaker timbre~\citep{wang2023valle, cho2025emosphere++}.

To support such control, existing TTS systems have explored various strategies.
Supervised models trained on labeled datasets~\citep{kim2021conditional, cho24_interspeech, leng2024prompttts} offer reliable conditioning but scale poorly due to costly annotations.
In contrast, reference-based zero-shot approaches~\citep{yourtts, wang2023valle, kim2023pflow, ji-etal-2024-mobilespeech, chen-etal-2024-f5tts} offer greater flexibility by conditioning on example utterances, but typically entangle style and timbre in a single reference, hindering independent control.

To overcome this limitation, recent studies have explored disentangled speech representations~\citep{choi2023nansy, ju2024naturalspeech}, leading to various TTS systems that exploit such factorization.
In particular, these systems employ disentangled attributes together with jointly pre-trained decoders to enable more interpretable control and, \emph{in principle}, the possibility of conditioning on style and timbre separately. Nevertheless, disentanglement is often imperfect in practice, and the decoders inherently bound synthesis quality, with no guarantee of robustness to unseen style-timbre combinations.

In this paper, we investigate how to effectively leverage disentangled speech representations to build TTS models that support separate control over style and timbre using distinct references.
To this end, we present FC-TTS, which leverages factorized speech representations but adds: (1) a two-stage spectrogram generation pipeline (timbre-conditioned blurry spectrogram, then prosody refinement) for robustness to unseen combinations; (2) a VQ-VAE-based style encoder that captures fine-grained and intra-utterance style variability; and (3) a conditioning-aware consistency loss that extends conventional regularization to multi-condition settings by enforcing joint coherence across timbre and style, providing more precise guidance for disentangled control.

In our experiments, we show that FC-TTS delivers competitive zero-shot TTS performance compared to state-of-the-art models that lack explicit support for separate control of style and timbre, as evaluated by objective metrics including UTMOS~\citep{saeki22c_utmos}, word error rate, and speaker similarity.
Furthermore, we conduct detailed evaluations on the RAVDESS dataset~\citep{livingstone2018ravdess}—a highly expressive emotional speech corpus—focusing on both timbre and prosody controllability through both objective and subjective evaluations.
Specifically, for timbre control, we compare FC-TTS with a factorized codec-based system~\citep{ju2024naturalspeech}, and for prosody control, we compare against a state-of-the-art zero-shot TTS model, F5-TTS~\citep{chen-etal-2024-f5tts}.
These evaluations highlight that FC-TTS not only maintains high synthesis quality but also enables precise and independent manipulation of style and timbre, which existing zero-shot systems do not explicitly support.

\section{Related Work}
\subsection{Conditional Text-to-Speech}
Conditional TTS incorporates auxiliary signals to control speaker or style attributes, enabling expressive and personalized speech.
A common approach is to use labeled datasets with speaker or style annotations~\citep{kim2021conditional, cho24_interspeech}, which support multi-condition models capable of controlling multiple attributes simultaneously~\citep{liu23t_promptstyle, liu-etal-2025-diffstyletts, kang23_zetspeech}.
More recently, prompt-guided methods~\citep{leng2024prompttts, ji-etal-2025-controlspeech} have emerged, where models interpret free-form descriptions of style or tone.
While these approaches offer disentangled controllability and flexibility, these approaches still rely heavily on annotation coverage and granularity, limiting scalability in open-domain or fine-grained control settings.

\subsection{Disentangled Speech Representation Learning}
A complementary direction focuses on learning disentangled speech representations that disentangle speech into interpretable components (\eg timbre, pitch, or linguistic content), opening promising directions for developing controllable TTS systems without labeled datasets.
NANSY++~\citep{choi2023nansy} leverages various information perturbation functions to isolate pitch, linguistic, and timbre features.
FACodec~\citep{ju2024naturalspeech} enforces strong information bottlenecks and uses supervision from phoneme and speaker labels to learn factorized representations.
LSCodec~\citep{guo2024lscodec} further improves timbre isolation via speaker perturbation and achieves strong results in voice conversion.
Despite progress, most systems are trained in autoencoding setups with paired inputs and targets, meaning they are primarily optimized for naturally co-occurring factors.
As a result, their ability to generalize to mismatched conditioning—such as combining timbre and prosody from different references—remains underexplored.

\subsection{Reference Speech-based Style and Timbre Control}
Recently, there have been studies to control the style and timbre of TTS models based on two reference samples.
For instance, IndexTTS 2~\citep{zhou2025indextts2} achieves this by relying on the disentanglement property of semantic codec representations derived from MaskGCT~\citep{wang2025maskgct}.
EmoSphere++~\citep{cho2025emosphere++} trains a dedicated emotion encoder and introduces a regularization loss to enforce orthogonality between emotion and timbre vectors.
Although promising, these methods rely on representations with only empirically observed disentanglement, or on pre-trained encoders that may discard other attributes such as accent.
In contrast, our approach builds on systematically factorized representations that preserve reconstruction quality while enabling explicit decomposition, providing a more reliable basis for separate style and timbre control.

\begin{figure*}[t]
  \centering
  \includegraphics[width=\textwidth]{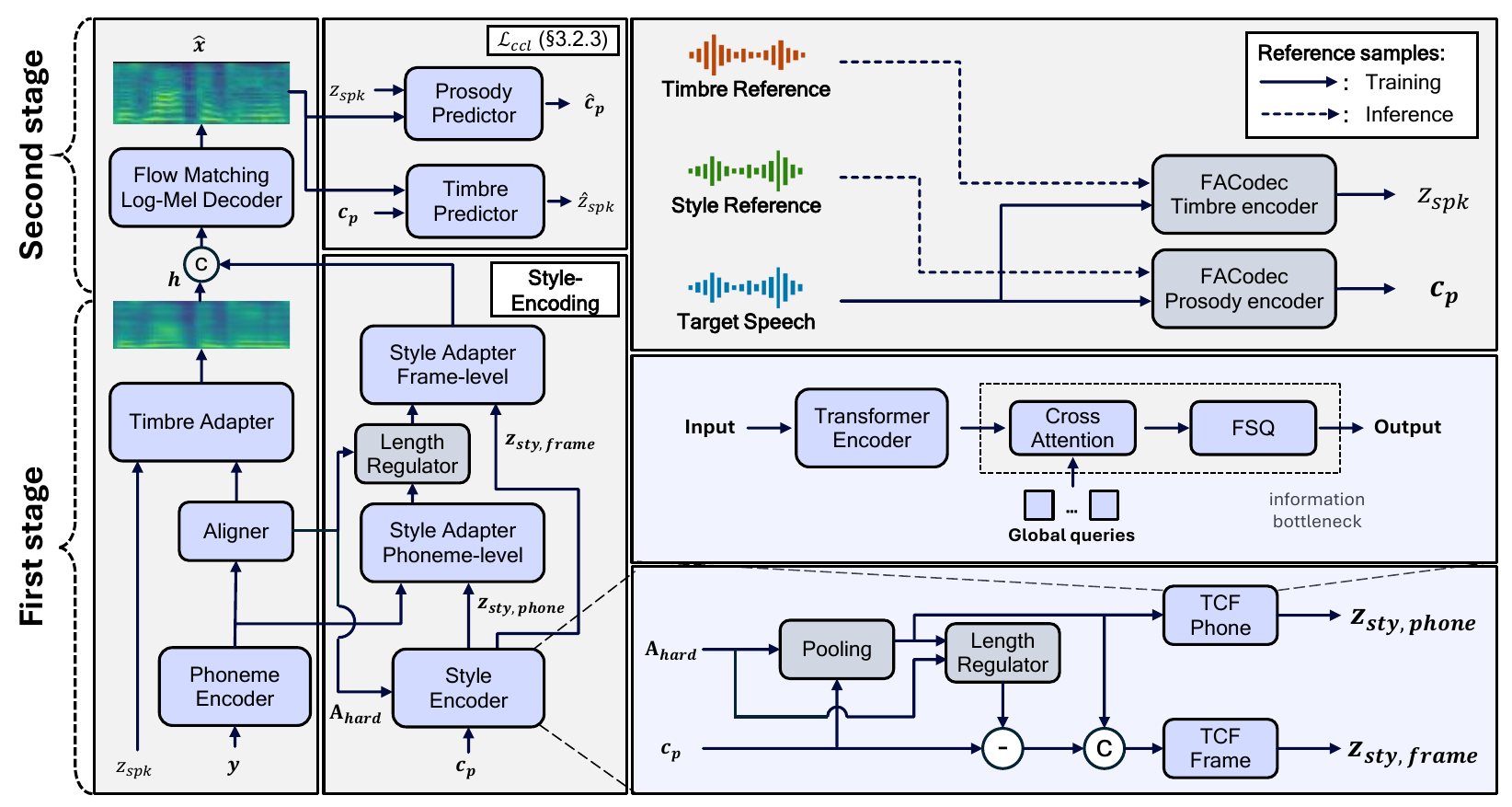}
  \caption{
  FC-TTS architecture. \textbf{First stage}: a phoneme sequence $\bm{y}$ is conditioned on timbre embedding $z_{\text{spk}}$ via the timbre adapter to generate a blurry log-mel spectrogram $\bm{h}$, anchoring timbre characteristics. \textbf{Second stage}: $\bm{h}$ is refined into a clean spectrogram $\hat{\bm{x}}$ by a flow-matching decoder conditioned on style embedding $\bm{z}_{\text{sty}}$, obtained from prosody tokens $\mathbf{c_p}$ through hierarchical TCF modules, imprinting prosodic characteristics. During training, target speech provides both timbre and style references; at inference, separate references enable disentangled attribute control. Blue modules indicate trainable components.
  }
  \label{fig:model}
\end{figure*}

\section{Methodology}
\subsection{Preliminaries}
\subsubsection{Factorized Speech Codec}
To build a timbre and style controllable TTS system, we adopt FACodec\footnote{\url{https://github.com/open-mmlab/Amphion/tree/main/models/codec/ns3_codec}}~\citep{ju2024naturalspeech}, a neural speech codec that decomposes speech into interpretable components while supporting faithful reconstruction.
It factorizes a speech signal into multiple disentangled streams of discrete tokens, each capturing a distinct speech attribute: prosody tokens $\mathbf{c_p}$, content tokens $\mathbf{c_c}$, and acoustic detail tokens $\mathbf{c_d}$.
Each stream is represented as $\mathbb{Z}^{N_* \times T}$, where $T$ is the number of time steps with residual quantization levels $N_p=1$, $N_c=2$, and $N_d=3$.
In addition, speaker timbre is captured as a continuous global embedding $z_{\text{spk}} \in \mathbb{R}^D$.
Notably, FC-TTS conditions exclusively on $z_{\text{spk}}$ and $\mathbf{c_p}$; the content tokens $\mathbf{c_c}$ and acoustic detail tokens $\mathbf{c_d}$ are deliberately excluded to prevent information leakage that would compromise independent control of the two pathways.

\subsubsection{Flow-matching TTS}
In conditional TTS, the goal is to generate a target speech representation $\bm{x} \in \mathbb{R}^{F \times T}$ (a log-mel spectrogram in this work) given a phoneme sequence $\bm{y} \in \mathbb{Z}^{L}$ and additional conditioning information $\bm{c}$ such as speaker timbre or speaking style.
To model this conditional generation process, we adopt the Conditional Flow-Matching (CFM) framework~\citep{lipman2023flow}, which defines a continuous-time transformation from a simple prior $p_1(\bm{x})$ (\eg isotropic Gaussian $\mathcal{N}(\mathbf{0}, \mathbf{I})$) to a target conditional distribution $p_0 = p(\bm{x}|\bm{y}, \bm{c})$.
To describe the progression from $p_1$ to $p_0$, CFM introduces a time-dependent flow $\phi_t \colon [0,1] \times \mathbb{R}^d \to \mathbb{R}^d$ that transports samples across time $t \in [0, 1]$, with marginal distribution $p_t(\bm{x})$ at each step. This flow is driven by a velocity field $v_t(\bm{x}) \colon [0,1] \times \mathbb{R}^d \to \mathbb{R}^d$, which specifies the instantaneous direction of motion at each point. Their relationship is governed by the following ordinary differential equation (ODE):
\begin{equation}
\frac{d}{dt} \phi_t(\bm{x}) = v_t(\phi_t(\bm{x})), \quad \phi_1(\bm{x}) = \bm{x}_1.
\end{equation}
Although the true $v$ is not available in practice, CFM approximates it by training  $u_\theta(\bm{x}, t, \bm{y}, \bm{c})$ with a conditional vector field $v_t(\bm{x}|\bm{x}_0)$.
Among possible flows, the straight-line optimal transport (OT) trajectory is known to be efficient, and the corresponding ground-truth velocity is given by:
\begin{equation}
v_t^{\text{OT}}(\bm{x}_t|\bm{x}_0) = \bm{x}_0 - \bm{x}_1,
\end{equation}
where $\bm{x}_t = (1 - t)\bm{x}_1 + t\bm{x}_0$. The model is then trained to align its predicted velocity $u_\theta$ with this OT velocity by minimizing the following loss:
\begin{equation}
\mathcal{L}_{\text{CFM}} = \mathbb{E}_{t, \bm{x}_0, \bm{x}_1} \left[ \left\| u_\theta(\bm{x}_t, t, \bm{y}, \bm{c}) - (\bm{x}_0 - \bm{x}_1) \right\|^2 \right].
\end{equation}

\subsection{FC-TTS}
Figure~\ref{fig:model} illustrates FC-TTS.
Although the timbre condition $z_{spk}$ and the style condition $\mathbf{c_p}$ are extracted from the same target during training, \emph{in principle}, the factorized codec allows these attributes to be controlled separately at inference using references from different utterances.
Building on this foundation, the following sections describe the architectural innovations that make such disentangled control practical.
Concretely, FC-TTS processes the two conditions sequentially across dedicated stages: a \emph{timbre stage} that anchors timbre characteristics via $z_{\text{spk}}$ to produce a blurry spectrogram, followed by a \emph{style stage} that imprints prosodic characteristics via $\mathbf{c_p}$ to refine it, ensuring each reference condition influences only its intended step.
Additional implementation details are provided in Section~\ref{subsec:setup} and Appendix~\ref{app:arc_detail}.

\subsubsection{Hierarchical Spectrogram Generation}
In preliminary experiments, we found that simply reusing the FACodec decoder, as in NaturalSpeech 3~\citep{ju2024naturalspeech}, was insufficient for independent timbre-prosody control. 
The main reason is that the imperfect disentanglement provides no guarantee of robust generation on unseen combinations.
To address this limitation, we propose a new model design that performs hierarchical log-mel spectrogram generation incorporating a jointly trained CFM speech decoder.
It first generates a blurry spectrogram $\bm{h}$ using timbre information, and subsequently refines it into a complete spectrogram $\bm{x}_0$ using style information through the CFM decoder.
These steps are trained jointly using mean-absolute-error (MAE) loss for the blurry spectrogram and CFM loss for the final output.
The MAE objective, defined as $\mathcal{L}_{\text{blur}} = \mathbb{E} \left[ \left\| \bm{h} - \bm{x}_0 \right\| \right]$, encourages over-smoothed outputs—a property we exploit to avoid the need for pre-generated blurry spectrograms.
Additionally, to prevent information leakage while maintaining consistent timbre and recording conditions, $z_{spk}$ is randomly replaced with another utterance from the same long audio file.
This two-stage design achieves functional separation: the timbre adapter injects $z_{\text{spk}}$ in the first stage to anchor timbre characteristics, while the style adapter subsequently applies $\mathbf{c_p}$ to imprint prosodic characteristics, ensuring each reference influences a dedicated processing pathway.

\subsubsection{VQ-VAE Style Encoding}
\begin{figure}[t]
  \centering
  \includegraphics[width=\columnwidth]{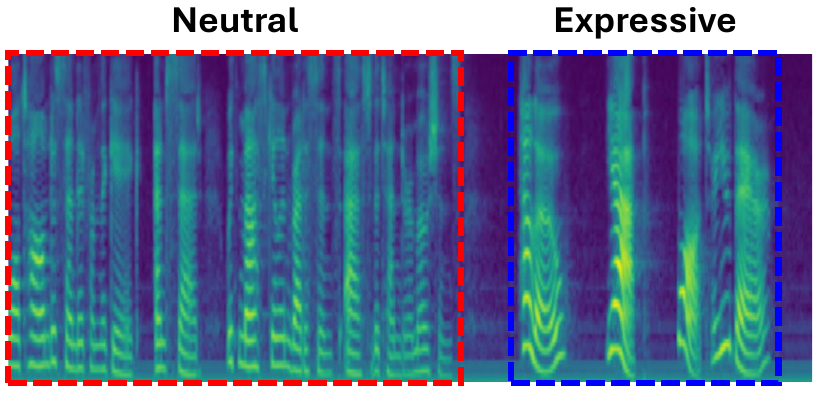}
  \caption{A speech sample from the Libriheavy dataset showing multiple speaking styles within a single utterance.
  }
  \label{fig:mixed}
\end{figure}

Recent zero-shot TTS models perform well in mimicking voice characteristics based on in-context learning (ICL), where part of the target speech is used as a prompt and the model generates the rest, assuming consistent timbre and style.
However, this assumption often fails in practice, as speaking style can vary even within a single utterance (Figure~\ref{fig:mixed}).
To address this, we condition the model on style representations extracted from the target speech during training, eliminating the need to assume style consistency.
However, this approach introduces a new challenge: the model may shortcut learning by copying surface-level acoustic features directly from the style reference, rather than capturing the intended higher-level prosodic patterns.
For this purpose, we propose a style encoder module called TCF, combining a Transformer encoder, Cross-attention, and a Finite scalar quantization layer (FSQ), which is instantiated twice within the main architecture to hierarchically model style representations at both the phoneme and frame levels~\citep{lei2023msstyletts}.
Each design component of TCF specifically targets one of these challenges:
\begin{itemize}
    \item \textbf{Prosody-only representation:} We exclusively use prosody tokens $\mathbf{c_p}$ from FACodec as the input to TCF, deliberately excluding content tokens $\mathbf{c_c}$ and acoustic detail tokens $\mathbf{c_d}$. This ensures that the style encoder captures rhythmic and intonational patterns without encoding unintended information.
    \item \textbf{Q-Former bottleneck~\citep{li2023blip}:} A fixed set of learned query tokens attends to the variable-length encoder outputs via cross-attention, compressing them into a fixed number of latent tokens. This bottleneck discards frame-level temporal details and forces the representation to retain only high-level stylistic structure, preventing the model from overfitting to the specific acoustic realization of the reference utterance.
    \item \textbf{Vector quantization~\citep{van2017vqvae}:} The continuous latent tokens produced by the Q-Former are further discretized using FSQ~\citep{mentzer2024finite}. Quantization acts as an information bottleneck that suppresses low-level acoustic residuals and encourages the encoder to commit to a discrete, semantically meaningful style code.
\end{itemize}
The detailed module architecture of TCF is provided in Appendix~\ref{app:arc_detail}.

\subsubsection{Conditional Consistency Loss}
To improve condition consistency in disentangled TTS, we present a conditional consistency loss (CCL) that extends prior regularization methods~\citep{xin21_interspeech, yourtts} to multi-condition settings. 
By enforcing joint coherence across prosody and speaker identity, CCL provides more reliable guidance for disentangled control.

\begin{figure}[t]
  \centering
  \includegraphics[width=\columnwidth]{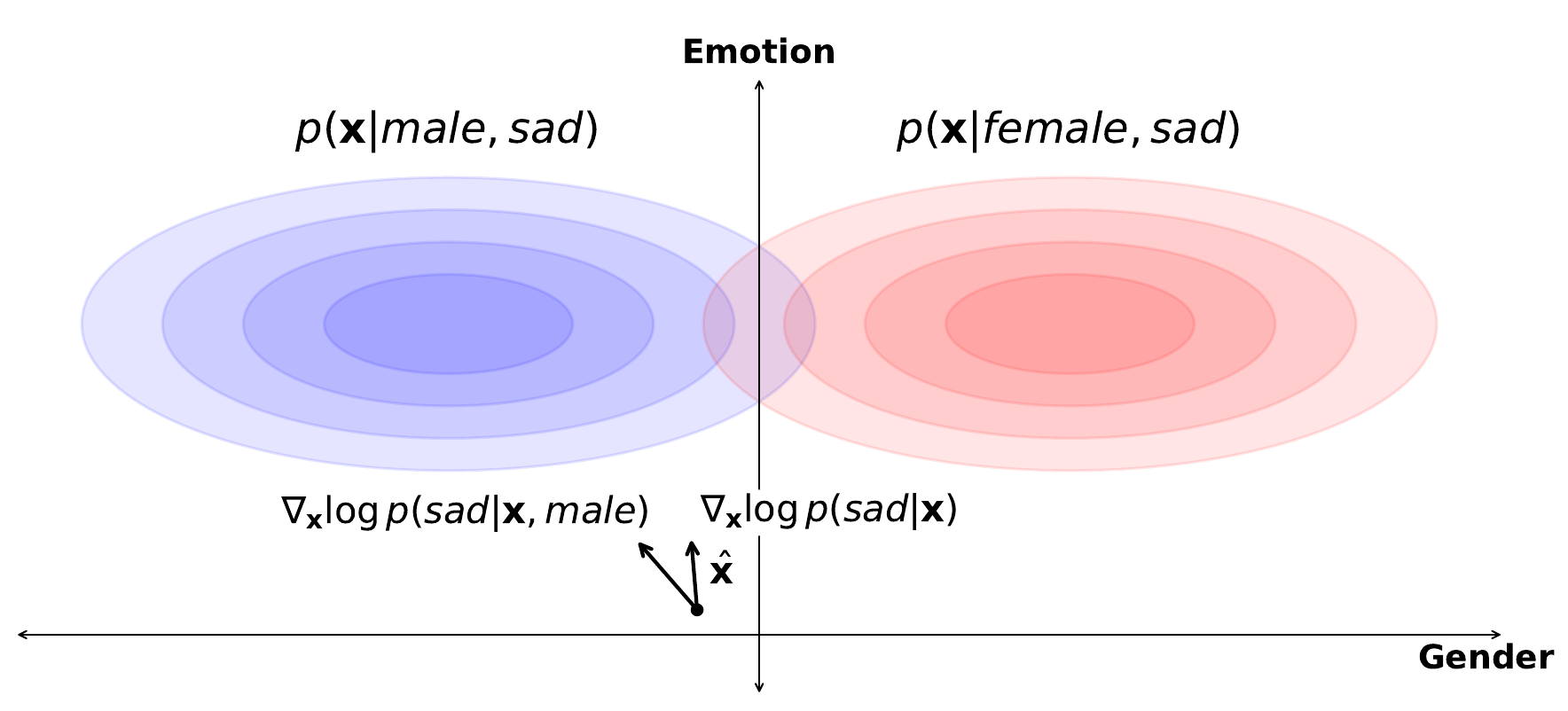}
  \caption{
  Gradients for CCL with multiple conditions.
  When a predictor takes only a log-mel spectrogram, the gradient $\nabla_{\bm{x}} \log p(\text{sad}|\bm{x})$ tends to point toward a region somewhere between the modes.
  In contrast, incorporating a known non-target attribute (\eg male) can sharpen the estimated posterior $p(\text{sad}|\bm{x}, \text{male})$, guiding spectrograms in more accurate directions, especially in early denoising steps when $\hat{\bm{x}}_0$ is only partially formed.
  }
  \vspace{-0.5cm}
  \label{fig:ccl_loss}
\end{figure}

We first re-parameterize the CFM objective~\citep{luo-etal-2025-wavefm} so that the FC-TTS decoder generates log-mel spectrograms directly, instead of a vector field.
Then, we train two attribute predictors on these spectrograms to predict the conditioning prosody token $\mathbf{c_p}$ and speaker embedding $z_{spk}$ respectively.
Importantly, each predictor also receives non-target conditioning signals, feeding $z_{spk}$ to the prosody predictor and $\mathbf{c_p}$ to the timbre predictor.
In Figure~\ref{fig:ccl_loss}, the effect of this cross-conditioning is illustrated through an example scenario involving two conditional labels: gender and emotion.

The CCL is defined as a weighted sum of two terms: a cross-entropy loss for prosody prediction and a negative cosine similarity for speaker embedding consistency:
\begin{align}
\mathcal{L}_{\text{CCL}} ={} & \lambda_{\text{ccl-pro}} \cdot \mathbb{E} \left[\text{CE}(\mathbf{c_p}, f(\hat{\bm{x}}, z_{\text{spk}})) \right] \notag \\
& - \lambda_{\text{ccl-spk}} \cdot \mathbb{E} \left[ \cos\left(z_{\text{spk}}, g(\hat{\bm{x}}, \mathbf{c_p})\right) \right],
\end{align}
where $\text{CE}(\cdot, \cdot)$ denotes the cross-entropy loss, $f(\cdot)$ denotes the prosody predictor, and $g(\cdot)$ denotes the speaker embedding predictor.

\section{Experiments}
This section presents a thorough evaluation of FC-TTS, focusing on its zero-shot TTS performance and the disentanglement of style and timbre. We first describe the experimental setup, including datasets, model architecture, and training-inference procedures. We then compare FC-TTS with state-of-the-art (SOTA) baselines and conduct ablation studies to assess the contribution of individual components.

\subsection{Experimental Setup}
\label{subsec:setup}
\subsubsection{Datasets}
For model training, we used the LibriHeavy dataset~\citep{kang2023libriheavy}, a large-scale audiobook corpus derived from LibriLight~\citep{librilight}. While it is not specifically curated for TTS, LibriHeavy offers substantial speaker and stylistic diversity, making it particularly useful for covering a wide range of speaking styles in zero-shot and expressive speech synthesis tasks.
Our primary evaluation was conducted on the LibriSpeech \texttt{test-clean} subset, a widely adopted benchmark for zero-shot TTS. This split is especially suitable for testing generalization, as it contains speakers unseen during training.
To further evaluate expressiveness and the disentanglement of style and timbre, we include experiments on the RAVDESS dataset~\citep{livingstone2018ryerson}, which contains emotionally rich speech across multiple speakers. This dataset serves as a complementary test set for evaluating FC-TTS's ability to independently model and control timbre and style in complex, high-variance vocal contexts.
Further details on datasets are provided in Appendix~\ref{app:data_detail}.

\subsubsection{Model Architecture}
The phoneme encoder is implemented using a transformer encoder architecture.
For alignment, we adopt the RAD-TTS aligner~\citep{shih2021rad} and train a CFM-based duration predictor using aligner-estimated durations.
The timbre adapter is a transformer encoder where the speaker embedding $z_{spk}$ conditions the layer normalization via adaptive layer normalization~\citep{perez2018film}.
The style adapter is a transformer decoder without causal masking, where style embeddings are provided through cross-attention layers.
The log-mel spectrogram decoder consists of DiT blocks~\citep{peebles2023scalable}.
The TCF modules consist of a transformer encoder, a cross-attention layer with global queries, and a finite scalar quantization layer.
Further architectural details are provided in Appendix~\ref{app:arc_detail}.

\subsubsection{Training \& Inference}
We train FC-TTS on LibriHeavy for 200k iterations with AdamW~\citep{loshchilov2018decoupled}, using a batch size of 64 and learning rate of 0.0002. 
During inference, FC-TTS first predicts durations using the duration predictor, with a number of function evaluations (NFEs) set to 8 and without classifier-free guidance.
For log-mel spectrogram synthesis, we use 32 NFEs with a classifier-free guidance scale of 4.0, and apply random conditioning dropout with a probability of 15\% during training.
Finally, the generated log-mel spectrograms are converted to waveforms using HiFi-GAN~\citep{kong2020hifi}.
Further details on training and inference are provided in Appendix~\ref{app:train_detail}.

\subsubsection{Baselines}
\label{subsubsec:baselines}
We benchmark FC-TTS against a diverse set of state-of-the-art TTS systems.
Specifically, we include \textbf{NaturalSpeech 3}~\citep{ju2024naturalspeech}, which represents one of the strongest FACodec-based models, \textbf{F5-TTS}~\citep{chen-etal-2024-f5tts}, which we retrain on LibriHeavy with a configuration comparable in size to ours, and reported results for \textbf{CLaM-TTS}~\citep{kim2024clamtts} and \textbf{DiTTo-TTS}~\citep{lee2025dittotts}, representing recent advances in zero-shot speech synthesis. 
In addition, we introduce a \textbf{FACodec-based voice conversion (VC)} system, which serves as the upper bound of FACodec-based TTS performance.
We adopt this setup for three reasons: (1) although several FACodec-based TTS models such as NaturalSpeech 3 have reported strong results, no official checkpoints are publicly available; (2) by directly reusing the ground-truth discrete tokens from the official FACodec encoder, this system approximates an idealized scenario in which a TTS model perfectly predicts codec tokens; and (3) by combining these tokens with an unmatched speaker embedding fed into the FACodec decoder, we simulate timbre controllability and assess the performance that a FACodec-based TTS system could achieve under such idealized conditions.

\subsubsection{Metrics}
We assess model performance using a comprehensive set of metrics: \textbf{UTMOS}\footnote{\url{https://huggingface.co/spaces/sarulab-speech/UTMOS-demo}}~\citep{saeki22c_utmos} for speech quality, \textbf{WER}\footnote{\url{https://huggingface.co/facebook/hubert-large-ls960-ft}} for pronunciation accuracy, \textbf{SPK}\footnote{\url{https://github.com/microsoft/UniSpeech/tree/main/downstreams/speaker_verification}} for speaker similarity, and \textbf{MCD} for prosodic and spectral similarity. These metrics collectively offer a multifaceted view of model performance across quality, intelligibility, and speaker consistency.
Although MCD is not a perfect measure of prosodic similarity, we adopt it because it has been widely used in prior works as a practical proxy~\citep{ju2024naturalspeech, yang2025measuring}.
Instead, we compute MCD with a fixed speaker to mitigate timbre confounding, so that the score primarily reflects prosodic differences.
Additionally, we conduct a \textbf{human ABX listening test} focused on timbre and prosody, providing a more direct perceptual measure of controllability.
To further supplement these evaluations with a more rigorous automated assessment of style similarity, we also employ an \textbf{AudioLLM-as-a-Judge} approach~\citep{chiang-etal-2025-audio, manku2025emergentttseval} using Gemini 2.5 Pro~\citep{comanici2025gemini} as the evaluator.
We measure two metrics: \emph{Win Ratio} via an ABX test that determines which of two generated samples better matches the reference style, and \emph{Style-MOS} that assigns a per-sample style similarity score (1--5).
Further details on metric definitions and evaluation tools are described in Appendix~\ref{app:eval_detail}.

\subsection{Results}
\subsubsection{Zero-Shot TTS Performance}
Table~\ref{tab:zstts} presents a comparative evaluation of FC-TTS against prior zero-shot TTS systems, focusing on naturalness and speaker similarity.
Following common practice, we evaluate on the LibriSpeech \texttt{test-clean} split and report values from original papers when available.
For evaluation, we use utterances from the \texttt{test-clean} subset that are 4 to 10 seconds long, with each text paired with a reference speech randomly sampled from the same speaker.
While FC-TTS does not achieve the absolute best scores across all metrics, it demonstrates competitive performance relative to SOTA TTS systems.
Importantly, unlike prior models, FC-TTS is explicitly designed for separate and controllable manipulation of timbre and style, a core capability not prioritized in existing TTS frameworks.
Although NaturalSpeech 3 also supports reference-based control, its disentanglement ability is limited, as we discuss in the next section.

We attribute the performance gap to a combination of deliberate design choices in favor of stronger disentanglement and structural constraints introduced by our generation pipeline.
First, we deliberately avoided using FACodec's content and detail tokens to prevent unintended information leakage; however, these tokens also carry useful cues that could enhance naturalness.
In particular, detail tokens often encode recording-environment characteristics, and without them the model tends to converge toward the average acoustic condition of LibriHeavy, a corpus not originally optimized for TTS production quality.
Second, our two-stage generation pipeline begins by producing a blurry spectrogram conditioned solely on timbre, which anchors the acoustic subspace that the subsequent style-based refinement stage must operate within.
This structural constraint limits the distribution the flow-matching decoder can reach compared to end-to-end systems that jointly optimize all attributes without an intermediate bottleneck.
While this design maximizes robustness to unseen style--timbre combinations, it may limit the ceiling of achievable naturalness relative to approaches such as NaturalSpeech 3 that impose no such constraint.
Third, due to the imperfect disentanglement of FACodec, residual timbre cues may still reside in the excluded tokens; handling such leakage more precisely remains an open challenge.
We view these as principled trade-offs that are necessary for achieving interpretable and independent controllability, and identify more robust disentanglement mechanisms—as well as codec-free alternatives—as key directions for future work.

\begin{table}[t]
\setlength{\tabcolsep}{1mm}
  \small
  \centering
  \begin{tabular}{lccccc}
    \toprule
    & UTMOS $\uparrow$ & WER $\downarrow$ & SPK $\uparrow$ & \#Param. $\downarrow$ \\
    \midrule
    Ground-truth                    & \nf{4.10} & \nf{2.07} & \nf{0.71} & - \\
    HiFi-GAN                        & \nf{3.70} & \nf{2.17} & \nf{0.64} & - \\
    \midrule
    NaturalSpeech 3                 &  \bf{4.30}  & \bf{1.81} & \bf{0.67} & 500M \\
    F5-TTS                          &      -      & \nf{2.42} & \nf{0.66} & 336M \\
    F5-TTS\textsuperscript{\dag}    &  \nf{4.03}  & \nf{3.30} & \bf{0.67} & 205M \\
    DiTTo-TTS                       &      -      & \nf{2.69} & \nf{0.60} & 508M \\
    CLaM-TTS                        &      -      & \nf{5.11} & \nf{0.50} & 584M \\
    FC-TTS (ours)                   &  \nf{4.22}  & \nf{1.88} & \nf{0.60} & 204M \\
    \bottomrule
  \end{tabular}
   \caption{Performance comparison of zero-shot TTS models. Reported values are taken from original papers if available. \textsuperscript{\dag} indicates models re-trained on Libriheavy with similar model size.}
  \label{tab:zstts}
\end{table}

\begin{table}[t]
\setlength{\tabcolsep}{1.0mm}
  \centering
  \small
  \begin{tabular}{lcccc}
    \toprule
    & UTMOS $\uparrow$ & SPK $\uparrow$ & WER $\downarrow$ & Win (\%) $\uparrow$ \\
    \midrule
    FACodec-VC        & \nf{3.19} & \nf{0.27} & \nf{8.40} &    10.7   \\
    Ours           & \bf{4.03} & \bf{0.48} & \bf{0.18} & \bf{66.1} \\
    \bottomrule
  \end{tabular}
  \caption{Objective metrics on the RAVDESS dataset for evaluating independent timbre control, a more challenging condition than standard TTS evaluation.}
  \label{tab:ravdess-vc}
\end{table}

\subsubsection{Timbre Controllability on RAVDESS}
In this experiment, we assess the model's robustness and its ability to accurately realize the intended timbre control when provided with prosodically richer reference samples from the RAVDESS dataset~\citep{livingstone2018ravdess}.
For each RAVDESS utterance used as a prosody reference, we synthesize speech by controlling the timbre according to two target speakers (one male and one female) selected from LibriSpeech \texttt{test-clean} subset.

In this experiment, we use the \textbf{FACodec-based voice conversion (VC)} system as the comparison model. 
As described in Section~\ref{subsubsec:baselines}, FACodec-VC directly reuses discrete tokens extracted from the FACodec encoder, approximating an oracle scenario where the codec tokens are perfectly predicted. 
Because there is no officially available FACodec-based TTS models (\eg NaturalSpeech 3), this configuration serves as a practical upper bound for the achievable performance of such systems.
By pairing these ground-truth FACodec tokens with distinct speaker embeddings, FACodec-VC enables a controlled assessment of whether FC-TTS can maintain timbre controllability independently of prosodic variations.

As summarized in Table~\ref{tab:ravdess-vc}, FACodec-based timbre control exhibits substantial degradation across all objective metrics, achieving a UTMOS of 3.19, a SPK of 0.27, and a WER of 8.40. 
In contrast, FC-TTS maintains strong performance, reaching a UTMOS of 4.03, a SPK of 0.48, and a WER of 0.18\footnote{The RAVDESS transcripts include only two short sentences — 
\emph{“Kids are talking by the door.”} and \emph{“Dogs are sitting by the door.”} — which explains the unusually low WER values.}, demonstrating its robustness under this mismatched setting.
Furthermore, we conduct a subjective ABX test in which listeners are presented with (A) the FACodec-VC output, (B) the FC-TTS output, and (X) a target reference, and asked which of A or B is closer to X in terms of \textit{timbre}. 
The resulting preference rate is reported as Win (\%) in Table~\ref{tab:ravdess-vc}. 
Consistent with the objective results, FC-TTS is preferred in 66.1\% of ABX trials, whereas FACodec-VC achieves only 10.7\%. 
Taken together, these results demonstrate that FC-TTS enables effective and independent timbre control even in prosodically complex scenarios, reinforcing the practical advantage of our disentangled design beyond conventional codec-based TTS.

\begin{table}[t]
\setlength{\tabcolsep}{1.0mm}
  \centering
  \small
  \begin{tabular}{lccccc}
    \toprule
    & UTMOS $\uparrow$ & SPK $\uparrow$ & WER $\downarrow$  & MCD $\downarrow$  & Win (\%) $\uparrow$ \\
    \midrule
    F5-TTS      & \nf{3.40} & \bf{0.57} & \nf{4.39} & \nf{3.43} &       8.9    \\
    Ours\textsuperscript{\dag}         & \bf{3.95} & \nf{0.47} & \bf{0.30} & \bf{3.21} & \bf{65.5} \\
    \bottomrule
  \end{tabular}
  \caption{Objective metrics on the RAVDESS dataset for evaluating independent style control. \textsuperscript{\dag} indicates models using separate reference inputs for timbre and prosody.}
  \label{tab:ravdess-pro}
\end{table}

\begin{table}[t]
\setlength{\tabcolsep}{1.0mm}
  \centering
  \small
  \begin{tabular}{lcc}
    \toprule
    & Win Ratio (\%) $\uparrow$ & Style-MOS $\uparrow$ \\
    \midrule
    F5-TTS   & \nf{8.3}  & \nf{1.50} \\
    Ours     & \bf{91.7} & \bf{3.92} \\
    \bottomrule
  \end{tabular}
  \caption{AudioLLM-as-a-Judge (Gemini 2.5 Pro) evaluation for style control on RAVDESS.}
  \label{tab:audiollm}
\end{table}

\subsubsection{Prosody Controllability on RAVDESS}
We next evaluate the prosody controllability of FC-TTS in comparison to F5-TTS using the RAVDESS dataset, as F5-TTS represents a SOTA zero-shot TTS system with publicly available source code.
For a fair comparison, we retrain F5-TTS on the LibriHeavy dataset using phoneme inputs and match its model size to that of FC-TTS.
Since F5-TTS does not support separate reference inputs for timbre and prosody, we provide a single reference speech during inference.
In contrast, FC-TTS leverages two distinct references: an expressive RAVDESS utterance for prosody and a neutral utterance from the same speaker for timbre.
This configuration imposes a more challenging inference condition but provides a clearer test of disentangled style control.

In addition to the automatic metrics used in the previous section, we further measure mel-cepstral distortion (MCD) between the expressive reference and the generated speech under matched text as an auxiliary indicator of prosody similarity. 
Although MCD is not a perfect proxy for prosodic attributes, we mitigate potential timbre confounding by fixing the timbre reference to a neutral utterance from the same speaker. 
We also complement these automatic evaluations with a subjective ABX prosody test, in which listeners judge which output—FC-TTS or F5-TTS—better matches the expressive reference in terms of prosodic pattern.

Results are summarized in Tables~\ref{tab:ravdess-pro} and~\ref{tab:audiollm}.
FC-TTS outperforms F5-TTS in UTMOS (3.95 vs.\ 3.40), WER (0.30 vs.\ 4.39), and MCD (3.21 vs.\ 3.43), while also being preferred in 65.5\% of ABX trials compared to only 8.9\% for F5-TTS.
These findings are further reinforced by the AudioLLM-as-a-Judge evaluation (Table~\ref{tab:audiollm}): Gemini 2.5 Pro assigns FC-TTS a Win Ratio of 91.7\% and a Style-MOS of 3.92, compared to only 8.3\% and 1.50 for F5-TTS, respectively, providing strong evidence that the style differences perceived by human listeners are also reliably captured by an automated large-scale evaluator.
Although FC-TTS yields a lower SPK score (0.47 vs.\ 0.57), this trade-off is consistent with the trends observed in Table~\ref{tab:zstts}; we note that maintaining perfect timbre consistency while achieving strong, disentangled style control from a separate reference—especially on highly expressive speech such as RAVDESS—represents an open research problem that we identify as a key direction for future work.
Overall, these results collectively demonstrate that FC-TTS achieves more accurate and disentangled prosody transfer without sacrificing intelligibility, even under stricter evaluation conditions.

\begin{table*}[ht!]
\setlength{\tabcolsep}{2.0mm}
  \centering
  \small
  \begin{tabular}{lccccccccc}
    \toprule
    & \multicolumn{4}{c}{Zero-shot TTS on LibriSpeech} & \multicolumn{4}{c}{Style Control on RAVDESS} \\
    \cmidrule(lr){2-5} \cmidrule(lr){6-9}
    & UTMOS $\uparrow$ & WER $\downarrow$ & SPK $\uparrow$ & MCD $\downarrow$ & UTMOS $\uparrow$ & WER $\downarrow$ & SPK $\uparrow$ & MCD $\downarrow$ \\
    \midrule
    FC-TTS                             & \nf{4.22} & \bf{1.88} & \bf{0.60} & \bf{5.60} & \nf{3.91} & \nf{0.30} & \bf{0.37} & \nf{3.33} \\
    $~~-$ two-stage generation              & \nf{4.15} & \nf{1.93} & \bf{0.60} & \nf{5.83} & \nf{3.57} & \nf{0.30} & \bf{0.37} & \bf{3.26} \\
    $~~-$ VQ-VAE style encoding                  & \bf{4.25} & \nf{2.00} & \nf{0.57} & \nf{5.62} & \bf{3.99} & \bf{0.25} & \nf{0.34} & \nf{3.47} \\
    $~~-$ conditioning in consistency loss             & \nf{4.21} & \nf{1.92} & \nf{0.59} & \nf{5.67} & \nf{3.79} & \nf{0.35} & \nf{0.36} & \nf{3.36} \\
    $~~-$ entire consistency loss   & \nf{3.95} & \nf{5.88} & \nf{0.48} & \nf{6.34} & \nf{3.70} & \nf{9.36} & \nf{0.21} & \nf{3.75} \\
    \bottomrule
  \end{tabular}
  \caption{Ablation study results. LibriSpeech use the same target data as reference for measuring MCD. RAVDESS uses the neutral speech as reference timbre and uses the emotional speech as reference style.}
  \label{tab:ablation1}
\end{table*}

\begin{figure}[ht]
  \centering
  \includegraphics[width=\columnwidth]{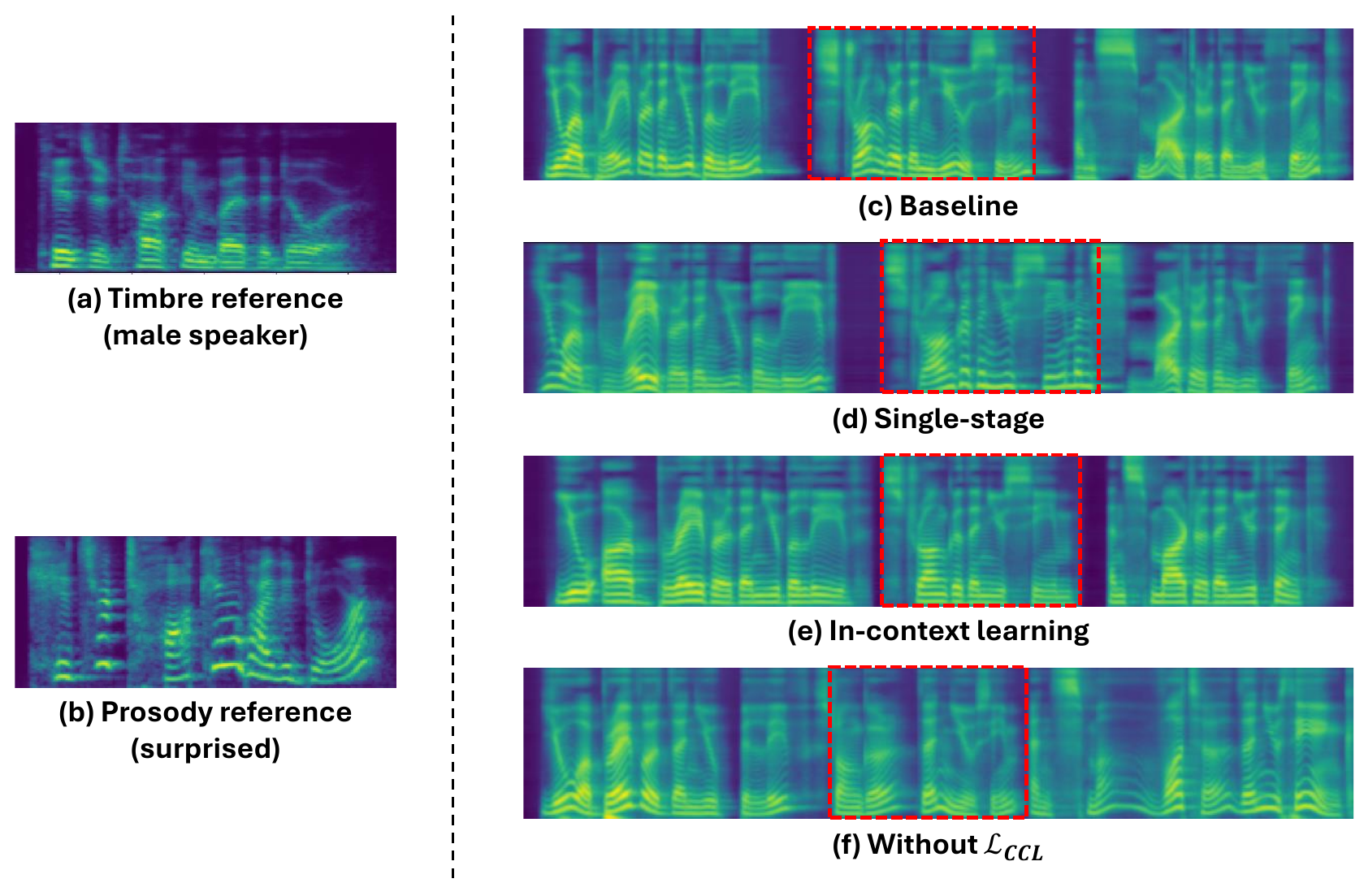}
  \caption{Log-mel spectrograms generated by each ablation variant (c-f), conditioned on the same timbre and prosody references (a, b) and using identical text input. The boxed regions highlight the segments corresponding to a specific phrase within the text input.}
  \label{fig:ablation}
\end{figure}

\subsubsection{Ablation Study}
We conduct an ablation study to analyze the contribution of three key components—two-stage generation, VQ-VAE style encoding, and conditional consistency loss—by systematically removing each from the full FC-TTS model. 
Table~\ref{tab:ablation1} summarizes quantitative results, and Figure~\ref{fig:ablation} illustrates representative spectrograms that highlight the distinct role of each component in shaping prosody and timbre.

\noindent\textbf{Two-stage generation.}
Removing this component degrades acoustic stability (UTMOS 4.15 vs.\ 4.22, MCD 5.83 vs.\ 5.60); Figure~\ref{fig:ablation}-(d) shows that the model over-reflects prosodic cues, yielding unstable acoustic patterns.
The two-stage design decouples coarse timbre shaping from fine prosody refinement, improving robustness to unseen style--timbre combinations.

\noindent\textbf{VQ-VAE style encoder.}
Removing this encoder yields a slight UTMOS gain (4.25 vs.\ 4.22) but notably weakens style control (SPK 0.57 vs.\ 0.60, MCD 3.47 vs.\ 3.33).
As visible in Figure~\ref{fig:ablation}-(e), the generated speech exhibits flattened F0 contours that fail to follow the target prosody.
Without the VQ-VAE encoder, the model falls back to ICL, which operates under an incorrect style assumption—treating the reference as stylistically uniform—and consequently fails to properly learn style reference following.

\noindent\textbf{Consistency loss.}
Removing only the cross-conditioning moderately worsens prosody alignment and intelligibility (MCD 3.36 vs.\ 3.33; WER 1.92 vs.\ 1.88).
Although the per-metric decrease is modest, this adds negligible training overhead and none at inference; the key design is that each predictor is conditioned on the other attribute (prosody predictor on $z_{\text{spk}}$, timbre predictor on $\mathbf{c_p}$), as visualized in Figure~\ref{fig:ccl_loss}, which we believe is a useful building block for future multi-attribute control.
Removing the \emph{entire} consistency loss causes catastrophic degradation—WER 5.88 (LibriSpeech) and 9.36 (RAVDESS) vs.\ 1.88 and 0.30—with inconsistent pitch and rhythm visible in Figure~\ref{fig:ablation}-(f), confirming that conditional supervision is the most indispensable component overall.

In summary, both the quantitative metrics in Table~\ref{tab:ablation1} and the spectrograms in Figure~\ref{fig:ablation} jointly demonstrate how each component contributes to prosody shaping and timbre stability. 
The results confirm that our proposed techniques—two-stage generation, VQ-VAE style encoding, and conditional consistency loss—play a central role in achieving robust and disentangled control.
For a deeper perceptual understanding of these effects, we strongly recommend listening to the audio samples on our demo page.\footnote{\url{https://qualcomm-ai-research.github.io/fc-tts}}

\section{Conclusion}
In this work, we presented FC-TTS, a zero-shot text-to-speech framework that enables independent control over timbre and speaking style using separate reference utterances.
While built upon factorized codec representations, FC-TTS introduces three complementary innovations that extend beyond the codec's limitations: (1) a two-stage spectrogram generation pipeline improving robustness to unseen timbre-style combinations, (2) a VQ-VAE-based hierarchical style encoder that captures fine-grained intra-utterance variation, and (3) a conditional consistency loss enforcing coherence across conditioning factors.
Extensive experiments on LibriSpeech and RAVDESS demonstrate that FC-TTS delivers competitive zero-shot quality and robust disentanglement, outperforming codec-based or single-reference baselines in both objective and perceptual evaluations.
Overall, FC-TTS provides a solid foundation for future research in diverse expressive TTS applications requiring disentangled and controllable speech synthesis.

\section{Limitations}
\paragraph{Language coverage and generalization.}
Our training and evaluation are currently limited to English datasets, which under-represent many languages, dialects, and accents. This restriction may limit the model's ability to generalize to diverse linguistic conditions. Extending FC-TTS to multilingual and cross-accent scenarios will be an important step toward assessing its robustness and adaptability in broader expressive TTS settings.

\paragraph{Dependence on codec representations.}
Although FC-TTS introduces architectural and training innovations beyond codec-based systems, it still relies on the disentanglement quality of FACodec representations. Consequently, its absolute synthesis quality may appear slightly below the strongest SOTA TTS models. Future work could explore more robust disentanglement mechanisms or codec-free formulations to further enhance fidelity and controllability.

\paragraph{Definition and interpretability of attributes.}
The conceptual boundary between timbre and style remains an open question. For instance, determining whether a ``husky voice'' should be treated as a stylistic or timbral attribute is not trivial. Clarifying such definitions and establishing quantitative metrics that can reliably assess these dimensions would further advance the interpretability, controllability, and scientific rigor of expressive TTS research.

\section{Ethical Considerations}
The zero-shot capability of FC-TTS, while offering substantial flexibility for expressive and personalized synthesis, also introduces potential misuse risks, particularly in the creation of deepfake or impersonated speech. By enabling highly realistic synthesis from minimal reference data, such models can reproduce a speaker's unique timbre and convey new emotional or stylistic content without their consent, raising concerns around privacy, identity theft, and misinformation.

These risks become more pronounced as FC-TTS allows independent style control while preserving a fixed timbre, which can be exploited to generate emotionally manipulated speech in another person's voice. To mitigate such misuse, future deployments of expressive TTS systems might consider architectures or interfaces that restrict style controllability while keeping timbre generation confined to authorized sources. We hope that future research will further explore technical and policy-oriented safeguards to balance creative expression with ethical responsibility in controllable TTS technologies.

\section*{Acknowledgments}

We are grateful to our colleagues for their support and encouragement throughout this research. In particular, we thank Guillaume Sautiere for insightful discussions and feedback on this work. The authors used AI writing assistance solely for minor language polishing and proofreading of the manuscript; it was not used to generate new content, ideas, or analyses.

\bibliography{custom}

\newpage
\appendix
\section{Architecture details}
\label{app:arc_detail}
This section elaborates on the FC-TTS architecture, expanding on components that were just briefly described from the main paper due to space constraints. Detailed hyperparameter configurations are listed in Table~\ref{tab:model-hyperparams}.

\begin{itemize}
\item \textbf{FACodec encoder:}
We utilize FACodec encoder to obtain disentangled speech representations: prosody tokens $\mathbf{c}_p$, content tokens $\mathbf{c}_c$, detail tokens $\mathbf{c}_d$, and the speaker embedding $z_{spk}$.
Since the temporal resolution of the discrete tokens differs from that of the log-mel spectrograms, we upsample the tokens to match the spectrogram's frame rate using a repetition-based method.
Additionally, we specifically utilize the tokens by first transforming them into embeddings via the codebook, followed by applying layer normalization without affine transformation to standardize the representations across their feature dimensions.

\item \textbf{Text encoder:}
The text encoder adopts a transformer-based architecture in which the traditional feed-forward network layers are substituted with one-dimensional convolutional layers. Positional information is encoded using rotary positional embeddings~\citep{SU2024127063_rope}. This configuration serves as the default transformer encoder architecture in this work.

\item \textbf{Aligner:}
This module aligns the output of the text encoder with the target log-mel spectrogram using an attention-based alignment search~\citep{shih2021rad}.
To accelerate alignment training, it utilizes two loss functions: $\mathcal{L}_{\text{forwardsum}}$, which promotes monotonic diagonal alignment via the connectionist temporal classification (CTC) algorithm~\citep{graves2006connectionist_ctc}; and $\mathcal{L}_{\text{bin}}$, which encourages the soft alignment $\mathcal{A}_{\text{soft}}$ calculated from the attention mechanism to be close to the binarized hard alignment $\mathcal{A}_{\text{hard}}$ calculated from the monotonic alignment search algorithm~\citep{NEURIPS2020_5c3b99e8_glowtts}.
During the initial training phase, the aligner uses $\mathcal{A}_{\text{soft}}$ to expand the text representations and starts to use $\mathcal{A}_{\text{hard}}$ after 10,000 iterations to reduce the training-inference gap. Operations requiring alignment—such as pooling and length regulation—also use $\mathcal{A}_{\text{hard}}$.

\begin{figure}[ht!]
  \centering
  \includegraphics[width=\columnwidth]{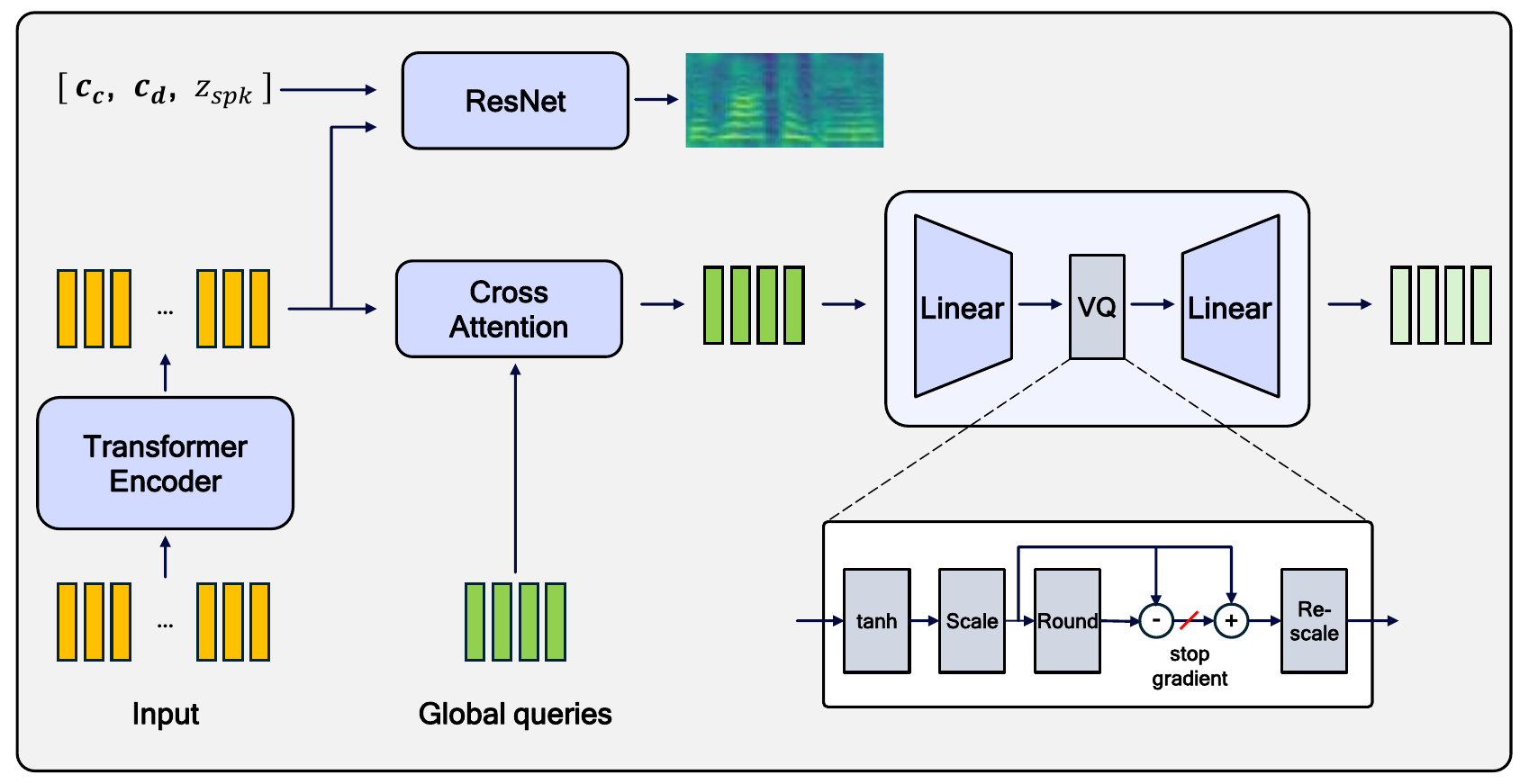}
  \caption{Detailed architecture of the TCF module. The input is first processed by a Transformer encoder, followed by compression using a Q-former-style cross-attention mechanism and a subsequence finite scalar quantization layer. To prevent representation collapse, a ResNet-based speech reconstruction module is jointly trained.}
  \label{fig:tcf}
\end{figure}

\item \textbf{Style encoder:}
Composed of two TCF modules operating at phoneme and frame levels respectively, this encoder models hierarchical style representations. It receives phoneme-level averaged prosody representations computed using $\mathcal{A}_{\text{hard}}$. For the frame-level input, it takes the concatenation of the re-expanded averaged prosody representations and the residual difference between the original and averaged representations to avoid encoding redundant information.

\item \textbf{Timbre adapter:}
It is built upon the transformer encoder with 1D CNNs. Also, the layer normalization is modified to adaptive layer normalization (AdaLN)~\citep{perez2018film}, allowing for the injection of the speaker embedding vector $z_{spk}$.

\item \textbf{Style adapter:}
It is built upon the transformer decoder with 1D CNNs. Also, the style embedding encoded by the style encoder is fed to the style adapter via the cross attention layer. Here, we do not use the causal masking typically used in the transformer decoder.

\item \textbf{Log-Mel decoder:}
The decoder adopts the DiT architecture~\citep{peebles2023scalable}, specifically the DiT Block with adaLN-Zero layers. Time embedding $t$ is injected into the adaLN layers.

\item \textbf{Prosody \& Timbre predictors:}
These attribute predictors are built using the transformer encoder architecture used in FACodec, allowing us to directly initialize them with the transformer modules from the pre-trained FACodec encoder.

\item \textbf{Duration predictor:}
This module follows the duration predictor used in MaskGCT~\citep{wang2025maskgct} and is trained using the CFM loss $\mathcal{L}_{\text{dur}}$ based on ICL. To encode duration references, it employs a transformer encoder block that processes a context prompt composed of log-scale durations, text representations, and phoneme-level averaged prosody representations extracted from a random segment corresponding to 0-30\% of the target. It adopts DiT blocks incorporated with cross-attention layers to effectively integrate the duration prompt.

\item \textbf{TCF module:}
The TCF module comprises a transformer encoder, a cross-attention mechanism, and a finite scalar quantization (FSQ) layer (see Figure~\ref{fig:tcf}).
During experimentation, we observed that the FSQ layer's latent representations tend to collapse into a single code.
To mitigate this, we introduce an auxiliary ResNet module trained to reconstruct the log-mel spectrogram using a mean absolute error loss $\mathcal{L}_{\text{mel-recon}}$.
The ResNet takes as input the output of the transformer encoder along with residual information necessary for spectrogram reconstruction.
\end{itemize}

\begin{table}[t!]
\centering
\begin{tabular}{@{}l c@{}}
\toprule
\multicolumn{2}{c}{\textbf{Training}} \\ \midrule
Learning Rate        & 0.0002       \\
Batch Size           & 64           \\
Iterations           & 200,000      \\
Optimizer            & AdamW        \\
Weight Decay         & 0.0          \\
Betas                & (0.8, 0.99)  \\
Warm-up              & 4,000        \\
Decay type           & exponential  \\
Decay factor         & 0.999875     \\
Decay step           & 200          \\
Gradient clip norm   & 10.0         \\
\bottomrule
\end{tabular}
\caption{Training hyperparameters}
\label{tab:train-hyperparams}
\end{table}

\section{Dataset details}
\label{app:data_detail}
In this work, we use three datasets:
\begin{itemize}
\item \textbf{Libriheavy:} 
A large-scale English read speech corpus consisting of 50,000 hours of labeled audio derived from LibriVox and Librilight.
It provides transcriptions with punctuation, casing, and textual context, enabling contextual ASR research.
The dataset is divided into training subsets of 500h (small), 5,000h (medium), and 50,000h (large) and we use all these subsets for training.
The dataset is available under an Apache-2.0 license at \url{https://github.com/k2-fsa/libriheavy}.

\item \textbf{LibriSpeech:}
A widely used 1,000-hour English ASR corpus based on public-domain LibriVox audiobooks and Project Gutenberg texts, distributed under the CC-BY 4.0 license. 
It includes training subsets of 100h, 360h, and 500h, and development and test sets of approximately 5 hours each (clean and other conditions).
In this work, we exclusively used the \textit{test-clean} subset (5.4 hours, 40 speakers: 20 male and 20 female) as our test set.
From this subset, we selected only utterances between 4 and 10 seconds in duration, resulting in approximately 1.2k audio samples.

\item \textbf{RAVDESS:} 
The Ryerson Audio-Visual Database of Emotional Speech and Song is a validated multimodal emotion dataset containing 7,356 recordings from 24 actors (12 male, 12 female). 
It includes audio-only, video-only, and audio-visual modalities across eight emotions for speech and six for song, each expressed at two intensity levels. 
The dataset is available under the CC BY-NC-SA 4.0 license.
\end{itemize}

\section{Training \& Inference details}
\label{app:train_detail}
\subsection{Training}
FC-TTS is trained on LibriHeavy dataset for 200,000 iterations using AdamW~\citep{loshchilov2018decoupled} optimizer, with a batch size of 64 and beta parameters set to (0.8, 0.99), on 8 NVIDIA V100 GPUs for 116 hours.
A learning rate of 0.0002 is employed, which undergoes a linear warmup over the initial 4,000 iterations before transitioning to exponential decay used in~\citet{kim2021conditional}.
Also, gradient norms are clipped at a maximum of 10.0 to stabilize training.

The overall training objective is defined by the following composite loss function:
$\mathcal{L}_{\text{total}}=
\lambda_{\text{CFM}} \cdot \mathcal{L}_{\text{CFM}}
+ \lambda_{\text{blur}} \cdot \mathcal{L}_{\text{blur}}
+ \lambda_{\text{ccl-pro}} \cdot \mathcal{L}_{\text{CE}}
+ \lambda_{\text{ccl-spk}} \cdot \mathcal{L}_{\text{cossim}}
+ \lambda_{\text{mel-recon}} \cdot \mathcal{L}_{\text{mel-recon}}
+ \lambda_{\text{forwardsum}} \cdot \mathcal{L}_{\text{forwardsum}}
+ \alpha \cdot \lambda_{\text{bin}} \cdot \mathcal{L}_{\text{bin}}
+ \alpha \cdot \lambda_{\text{dur}} \cdot \mathcal{L}_{\text{dur}}.
$
Here, each $\lambda_*$ term controls the contribution of its corresponding loss component.
The coefficients are set as follows: $\lambda_{\text{CFM}}=5.0,~\lambda_{\text{blur}}=1.0,~\lambda_{\text{ccl-pro}}=0.2,~\lambda_{\text{ccl-spk}}=0.5,~\lambda_{\text{mel-recon}}=1.0,~\lambda_{\text{dur}}=1.0,~\lambda_{\text{forwardsum}}=0.1$, and $\lambda_{\text{bin}}=0.1$.
Additionally, the coefficient $\alpha$ is linearly warmed up over the first 10,000 iterations to allow time for the aligner to estimate accurate alignments.

\subsection{Inference}
Inference proceeds in two stages: duration prediction and spectrogram generation. Duration prediction is performed as the first stage of inference, using a fixed number of function evaluations (NFEs), set to 8. This stage does not incorporate classifier-free guidance.
For log-mel spectrogram synthesis, we use 32 NFEs with a classifier-free guidance scale of 4.0. To enable classifier-free guidance, we randomly drop the conditioning inputs during training with a probability of 15\%.
Lastly, the generated log-mel spectrograms are transformed into 22kHz waveforms using a pre-trained HiFi-GAN~\citep{kong2020hifi} vocoder.
To match the input requirements of HiFi-GAN, we use LibriHeavy speech samples that have been upsampled to 22kHz.
Despite the waveform synthesis operating at 22kHz, the HiFi-GAN vocoder is trained with log-mel features configured with an \texttt{f\_max} of 8000 Hz, which enables compatibility between features extracted from 16kHz audio and the final waveform synthesis at 22kHz.
Further hyperparameter settings are listed in Table~\ref{tab:train-hyperparams}.

\section{Evaluation details}
\label{app:eval_detail}
This section provides supplementary details on the evaluation setup and methodology.
We compare the performance of FC-TTS against a diverse set of baseline models to ensure a comprehensive evaluation:
\subsection{Baselines}
\begin{itemize}
  \item \textbf{NaturalSpeech 3 (NS3)}: Selected as the primary baseline due to its strong performance and architectural similarity to FC-TTS, notably its use of FACodec. When we asses its separate controllability of timbre and style, we instead evaluate the voice conversion capability of FACodec, which serves as an upper bound for NS3's performance in this context.
  \item \textbf{F5-TTS}~\citep{chen-etal-2024-f5tts}: An in-context learning-based state-of-the-art zero-shot TTS model. We selected this model because its official source code is publicly available under MIT License, allowing us to retrain it under identical conditions and thereby ensure both fairness in evaluation.
  \item \textbf{CLaM-TTS \& DiTTo-TTS}~\citep{kim2024clamtts, lee2025dittotts}: These models are also included to broaden the comparison, using reported metrics evaluated on the same test set and scoring protocol.
  \item \textbf{FACodec}~\citep{ju2024naturalspeech}: This is used a voice conversion system to approximate the oracle of the FACodec-based TTS models. Based on the official checkpoint, which is available under MIT License, we first extract ground-truth FACodec tokens from the encoder and reconstruct it with unmatched timbre embedding using FACodec decoder.
\end{itemize}
\subsection{Metrics}
Also, we assess model performance using a diverse set of objective metrics as follows:
\begin{itemize}
  \item \textbf{UTMOS}\footnote{\url{https://huggingface.co/spaces/sarulab-speech/UTMOS-demo}}~\citep{saeki22c_utmos} (MIT License): A neural network-based metric trained to predict human mean opinion scores (MOS) for speech quality. It provides a reliable proxy for perceptual naturalness and fluency, and is widely adopted for evaluating TTS systems.
  \item \textbf{WER (Word Error Rate)}: Computed using a pre-trained HuBERT-based automatic speech recognition model\footnote{\url{https://huggingface.co/facebook/hubert-large-ls960-ft}} (Apache 2.0 License). WER quantifies intelligibility by comparing the transcribed output of generated speech against ground-truth text, with lower values indicating better pronunciation accuracy.
  \item \textbf{SPK}: A metric for evaluating speaker similarity, computed as the cosine similarity between embeddings extracted using a WavLM-TDCNN-based speaker verification model\footnote{\url{https://github.com/microsoft/UniSpeech/tree/main/downstreams/speaker_verification}} (CC BY-SA 3.0 License). These embeddings are obtained from both reference and synthesized speech samples. Higher scores indicate greater similarity in speaker characteristics.
  \item \textbf{MCD (Mel Cepstral Distortion)}: MCD calculates the Euclidean distance between mel-cepstral coefficients of two speech signals. These coefficients capture the spectral envelope of speech.
  In our evaluation, we use MCD for assessing prosodic similarity between the reference and generated utterances, which contain identical linguistic content and are spoken by the same speaker.
\end{itemize}

\subsection{ABX test}
The human listening evaluation was conducted with internal company employees serving as participants. To ensure fairness and reliability, a set of test samples was randomly selected from the evaluation dataset. For each sample, corresponding speech outputs were generated by both the baseline system and our proposed system. During the test, participants were presented with pairs of audio clips in randomized order—one from each system—without any indication of their origin. They were asked to identify which sample more closely matched the reference in terms of naturalness and intelligibility.
This procedure was designed to minimize potential bias and to ensure the validity and reproducibility of the evaluation results.

\subsection{AudioLLM-as-a-Judge Evaluation}
\label{app:audiollm}
We use Gemini 2.5 Pro~\citep{comanici2025gemini} as the evaluator model and measure two complementary metrics.
\textbf{Win Ratio} is obtained via pairwise comparison: the model is given a human reference and two generated samples and judges which better matches the reference speaking style; the aggregate preference rate over all test pairs constitutes the Win Ratio.
\textbf{Style-MOS} is obtained via independent per-sample scoring: the model rates how well a single generated sample reflects the speaking style of a given reference on a 1--5 scale, and the mean over all pairs constitutes the Style-MOS.

\begin{table*}[t!]
\centering
\begin{tabular}{@{}c l c@{}}
\toprule
\textbf{Module} & \textbf{Hyperparameter} & \textbf{Value} \\ \midrule

\multirow{5}{*}{Text encoder / Duration prompt encoder} 
  & Hidden dimension     & 192 \\
  & Layers               & 6 \\
  & Heads                & 2 \\
  & FFN dimension        & 768 \\
  & Conv kernel size     & 3 \\ \midrule

\multirow{5}{*}{Duration predictor} 
  & Hidden dimension     & 192 \\
  & Layers               & 6 \\
  & Heads                & 4 \\
  & FFN dimension        & 768 \\
  & Conv kernel size     & 3 \\ \midrule

\multirow{5}{*}{Timbre adapter / Style adapter-phone / Style adapter-frame} 
  & Hidden dimension     & 192 \\
  & Layers               & 4 \\
  & Heads                & 4 \\
  & FFN dimension        & 768 \\
  & Conv kernel size     & 9 / 5 / 9 \\ \midrule

\multirow{5}{*}{Log-mel decoder} 
  & Hidden dimension     & 384 \\
  & Layers               & 12 \\
  & Heads                & 4 \\
  & FFN dimension        & 768 \\
  & Conv kernel size     & 5 \\ \midrule

\multirow{5}{*}{Prosody / Timbre predictor} 
  & Hidden dimension     & 256 \\
  & Layers               & 4 \\
  & Heads                & 4 \\
  & FFN dimension        & 1024 \\
  & Conv kernel size     & 5 \\ \midrule

\multirow{10}{*}{TCF-phone / TCF-frame} 
  & Hidden dimension           & 384 \\
  & Layers                     & 6 \\
  & Heads                      & 4 \\
  & Conv kernel size           & 5 \\
  & Queries                    & 4 \\
  & FSQ latent dimension       & 6 \\
  & FSQ latent bins            & [5, 5, 5, 5, 5, 5] \\
  & ResNet Blocks              & 3 \\
  & ResNet Conv Layers         & 2 \\
  & ResNet Conv kernel size    & [[9, 1], [9, 1], [9, 1]] \\
  & ResNet Conv dilations      & [[1, 1], [2, 1], [4, 1]] \\

\bottomrule
\end{tabular}
\caption{Grouped hyperparameters for different modules. Modules with similar configurations are merged into single group, and differing values are separated by slashes (e.g., “Timbre adapter / Style adapter-phone / Style adapter-frame”) to reduce redundancy and improve readability.}
\label{tab:model-hyperparams}
\end{table*}

\begin{figure*}[ht!]
  \centering
  \includegraphics[width=0.7\textwidth]{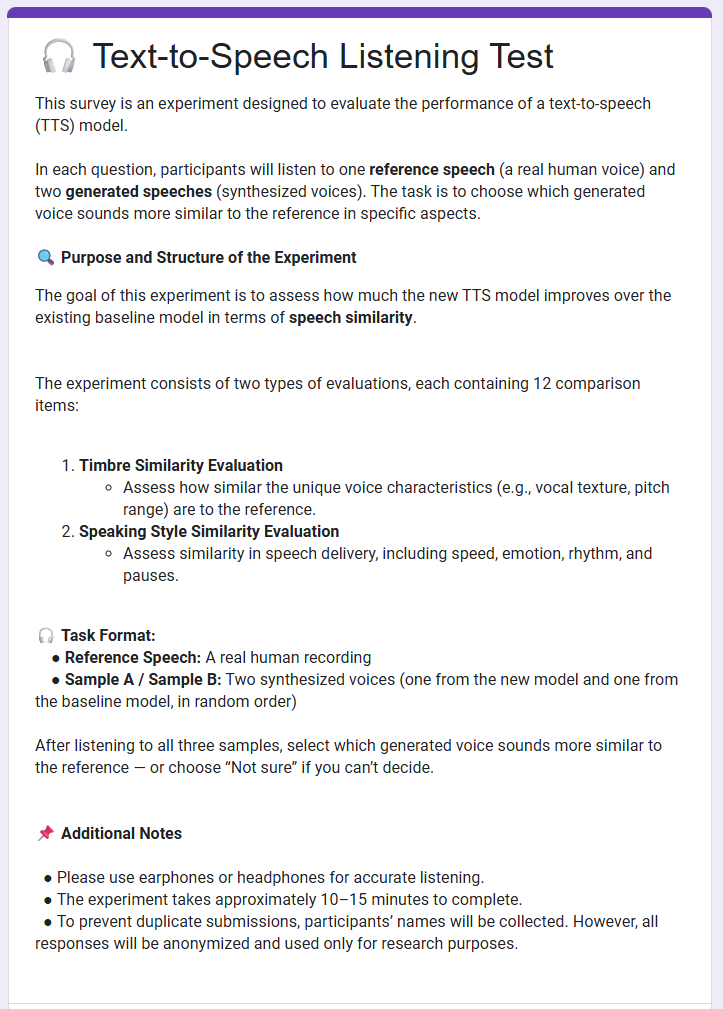}
  \caption{Main instructions provided to participants in the ABX test.}
\end{figure*}

\begin{figure*}[ht!]
  \centering
  \includegraphics[width=0.7\textwidth]{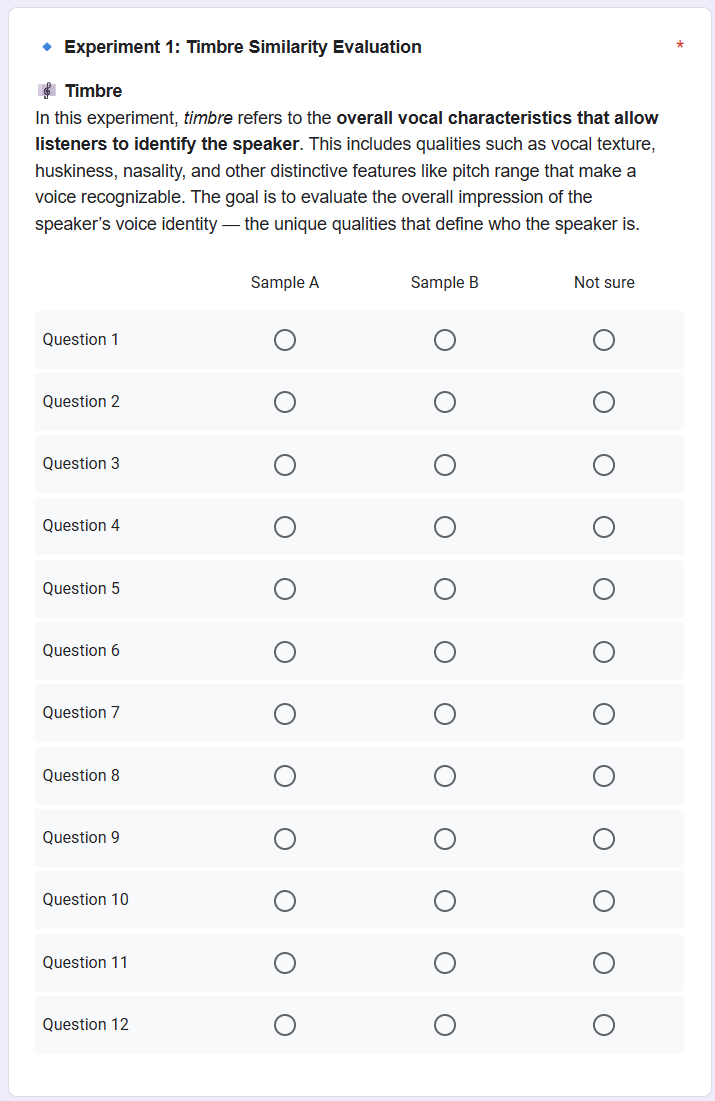}
  \caption{Instructions for the timbre controllability evaluation.}
\end{figure*}

\begin{figure*}[ht!]
  \centering
  \includegraphics[width=0.7\textwidth]{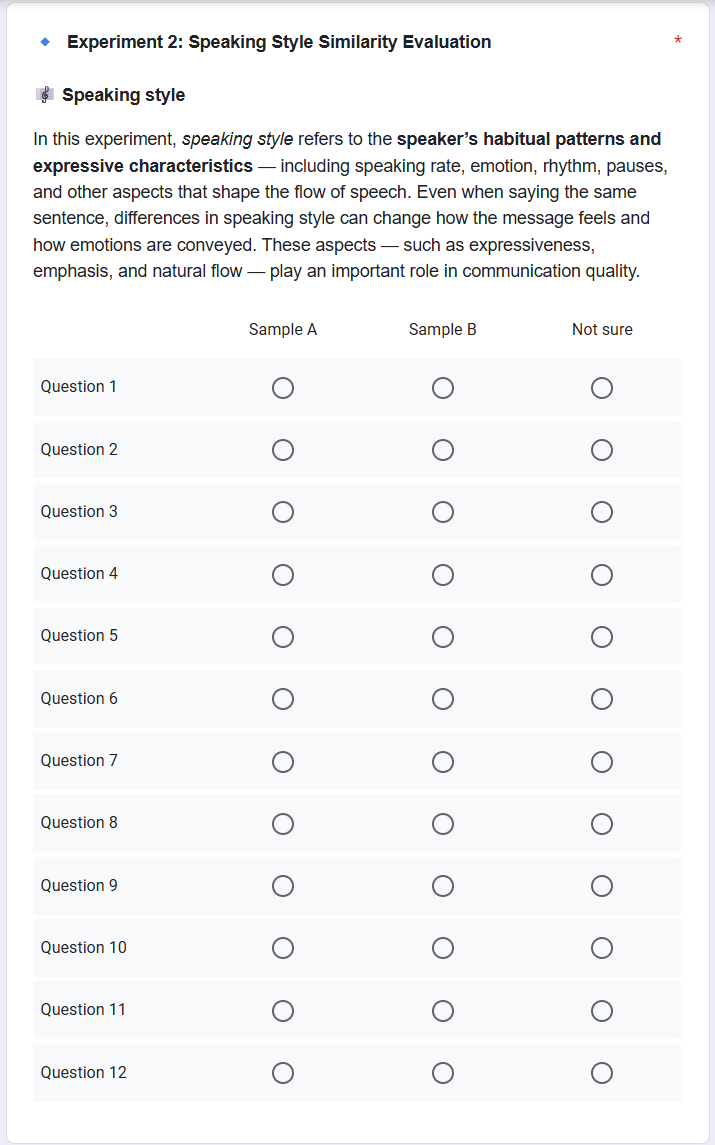}
  \caption{Instructions for the style controllability evaluation.}
\end{figure*}

\end{document}